\documentclass[preprintnumbers,showpacs,amsmath,amssymb,floatfix,prd,onecolumn,superscriptaddress,nofootinbib,12pt]{revtex4}
\usepackage{amssymb}
\usepackage{graphicx}
\usepackage{epsfig}
\usepackage{bm}
\usepackage{mathrsfs}
\usepackage{amsfonts}
\usepackage{epstopdf}
\usepackage{bm}
\usepackage{amsthm}
\usepackage{multirow}
\usepackage{subfigure}
\usepackage{amsmath}
\usepackage{textcomp}
\usepackage{fancyhdr}
\begin{document}
	\title{Entropies and The First Laws of Black Hole Thermodynamics in Einstein-aether-Maxwell Theory}

	\author{Hai-Feng \surname{Ding}}
	\email[]{haifeng1116@qq.com}
	
	\author{Xiang-Hua \surname{Zhai}}
	\email[]{zhaixh@shnu.edu.cn}

	\affiliation{Division of Mathematical and Theoretical Physics, Shanghai Normal University, 100 Guilin Road, Shanghai 200234, China}

\begin{abstract}
	Using the solution phase space method, we investigate the thermodynamics of black holes in Einstein-aether-Maxwell theory, for which the traditional Wald method (covariant phase space method) fails. We show the first laws of thermodynamics and definitive entropy expressions at both Killing and universal horizons for some examples of exact black hole solutions, including 3-dimensional static charged quasi-BTZ black hole, two 4-dimensional static charged black holes and 3-dimensional rotating solution. At Killing horizons the entropies are exactly one quarter of the horizon area, but at universal horizons of 3-dimensional black holes, the entropies have a corrected term in addition to the one proportional to the horizon area.
\end{abstract}

	\maketitle
	
	
	\maketitle

	\newpage
	\section{Introduction}
	As a family of modified gravitational theories, the Lorentz violating (LV) gravitational theories \cite{Mattingly:2005re}, including Hořava-Lifshitz theory \cite{Horava:2009uw}, ghost condensation \cite{ArkaniHamed:2003uy}, warped brane world and Einstein-aether theory \cite{Jacobson:2000xp,Eling:2004dk,Jacobson:2008aj}, have attracted increasing interests. Einstein-aether theory is a diffeomorphism-invariant theory of gravity that violates the local Lorentz invariance, in which a unit time-like dynamical vector field (the so-called aether field) $u^{\mu}$ was introduced to couple to the Einstein's general relativity (GR) \cite{Jacobson:2000xp}. The existence of the aether field defines a preferred frame, which leads to the local Lorentz invariance violation and yields some interesting consequences: matter fields can travel faster than the speed of light \cite{Jacobson:2000gw}, the causality of the theory is quite different from that of GR, and new gravitational wave polarizations can propagate at different speeds \cite{Jacobson:2004ts}. In the LV theories, the propagating speed of the particles can be arbitrarily large, so the corresponding ``light cones” can be completely flat and the causality of the theory is more like that of the Newtonian theory \cite{Greenwald:2011ca,Wang:2017brl}. As a result, an important feature of the LV theories is that the event horizon of the black hole is the universal horizon (UH) , instead of the Killing horizon (KH) in the Lorentz symmetric theories of gravity. The UH is a hypersurface acting as a causal boundary and a one-way membrane, which can trap the modes of arbitrarily high velocity. The UH also defines the black hole region in space-time. 
	
	So far, several exact black hole solutions with UH have been found in LV gravitational theories, including four-dimensional static asymptotically flat solutions of aether theory and Hořava gravity  \cite{Blas:2011ni,Barausse:2011pu,Berglund:2012bu}, four-dimensional static asymptotically (anti-) de sitter (Ads) solutions \cite{Bhattacharyya:2014kta}, three-dimensional fully rotating solution of infrared (IR) limit of Hořava gravity \cite{Sotiriou:2014gna} in the branch $c_{14}=0$, the black brane solutions with asymptotically Lifshitz property \cite{Janiszewski:2014iaa}, charged static solutions in 3 and 4 dimensional \cite{Ding:2016wcf,Ding:2015kba}, and the generalization to D dimensional static charged solutions \cite{Lin:2017cmn,Ding:2018whp}.
	
	Recently, by using the tunneling method to study the corresponding Hawking radiation at UH \cite{Berglund:2012fk,Ding:2015fyx,Cropp:2013sea,Michel:2015rsa}, it was shown that the black hole radiates at the UH like a blackbody and has a thermodynamic interpretation. By using the Clausius relations, the first law of Einstein-aether black hole was studied in \cite{Liberati:2017vse}. The Smarr integral formula for Einstein-aether theory was given in \cite{Ding:2016wcf,Ding:2015kba,Pacilio:2017emh,Ho:2017eot}. However, it is still an open question to generalize the first law of black hole thermodynamics to charged and/or rotating black hole solutions in the LV gravitational theories \cite{Foster:2005fr,Mohd:2013zca,Ding:2016wcf,Ding:2015kba,Liberati:2017vse,Pacilio:2017emh,Ho:2017eot}. The main reason is that traditional thermodynamic methods are invalid for deriving the first law for this class of black hole solutions. The Wald’s formulation \cite{Lee:1990nz,Wald:1993nt,Iyer:1994ys}, as a well known traditional thermodynamic method, has been applied to Einstein-aether theory\cite{Foster:2005fr,Mohd:2013zca}, and the expressions of the total energy, momentum, and angular momentum have been obtained. But as was mentioned in \cite{Foster:2005fr,Mohd:2013zca}, because of the divergence of the aether field on the KH bifurcation surface, this formalism fails for this theory to derive the first law of black hole thermodynamics and to give the definitive expression for black hole entropy.  Another reason is that the expressions of horizon location $r_{{}_{\mathrm{KH}}}$ are too complex, so it is hard to obtain the first law and the definitive entropy expressions from an explicit black hole solution.
	
	Based on the covariant phase space method (CPSM) \cite{Lee:1990nz,Wald:1993nt,Iyer:1994ys} (i.e. the Wald’s method), K. Hajian \textit{et al} developed the solution phase space method (SPSM) \cite{Hajian:2015xlp,Hajian:2016kxx,Ghodrati:2016vvf} for calculating the conserved charges (mass, angular momentum, electric charge as well as entropy) associated with the ``exact symmetry’’ of black hole solutions. With the SPSM, the conserved charges can be calculated by an integration over an almost arbitrary smooth codimension-2 surface surrounding the singularity of the black hole (not necessarily restricted to the horizon bifurcation surface on which CPSM depends), independent of any specific horizon or asymptotics. The entropy is also interpreted as a conserved charge associated with an exact symmetry, which would be a linear combination of generators of stationarity, axial isometry, and global gauge transformation. The linear combination coefficients are determined by the choice of the horizon and are exactly the coefficients appearing in the first laws. Therefore, entropy and the first law can be derived for any chosen horizon of explicit black hole solution, and the singularity of aether field at KH will not affect the final results, as the integration is not restricted to KH. The problem presented in \cite{Foster:2005fr,Mohd:2013zca} can thus be resolved by using the SPSM.
	
	In this paper, we will use the SPSM to investigate the thermodynamics of black holes in Einstein-aether-Maxwell Theory. We will give the first laws of black hole thermodynamics and definitive entropy expressions at both KH and UH for several exact black hole solutions, including 3-dimensional static charged quasi-BTZ black hole, two 4-dimensional static charged black holes, and 3-dimensional rotating solution.
	
	The rest of the paper is organized as follows. In Sec. II we briefly review the SPSM. In Sec. III the Noether-Wald charge and surface charge are derived for Einstein-aether-Maxwell theory. In Sec. IV, we use the SPSM to give the first laws and definitive entropy expressions for some exact black hole solutions in Einstein-aether theory. Section V is dedicated to main conclusions and remarks.
	
\section{Solution phase space method}
	The SPSM \cite{Hajian:2015xlp,Hajian:2016kxx,Ghodrati:2016vvf} is based on the CPSM \cite{Lee:1990nz,Wald:1993nt,Iyer:1994ys}, so we will first give a brief review of CPSM to obtain the conserved charges for generic gauge theories.
	
	\textbf{Covariant phase space:} A phase space is a manifold $\mathcal{M}$ equipped with a symplectic 2-form $\Omega$. Let us consider an $n$-dimensional generally covariant gravitational theory described by Lagrangian $n$-form  $\mathbf{L}$. To construct $\Omega$, we take the variation of  $\mathbf{L}$ as
	\begin{equation}\label{key}
	\delta \mathbf{L}(\Phi)=\mathbf{E}_{\Phi} \delta \Phi+\mathrm{d} \mathbf{\Theta}(\delta \Phi, \Phi) ,
	\end{equation}
	where $\Phi$ is used to denote collectively all the dynamical fields and $\mathbf{E}_{\Phi}=0$ gives the equations of motion (EOM). $\mathbf{\Theta}$ is an $n-1$-form called the Lee-Wald symplectic potential.
	The Lee-Wald symplectic form \cite{Lee:1990nz} is defined as
	\begin{equation}\label{key}
	\Omega \left(\delta_{1} \Phi, \delta_{2} \Phi, \Phi\right)=\int_{\Sigma} \bm{\omega} \left(\delta_{1} \Phi, \delta_{2} \Phi, \Phi\right) ,
	\end{equation}
	where
	\begin{equation}\label{key}
	\bm{\omega} \left(\delta_{1} \Phi, \delta_{2} \Phi, \Phi \right) \equiv \delta_{1} \mathbf{\Theta} \left(\delta_{2} \Phi, \Phi\right)-\delta_{2} \mathbf{\Theta} \left(\delta_{1} \Phi, \Phi\right)
	\end{equation}
	is the symplectic current form, and $\Sigma$ is a smooth codimension-1 Cauchy surface.
	
	When $\Phi$ solves the EOM $\mathbf{E}_{\Phi}=0$ and $\delta \Phi$ solves the linearized EOM $\delta \mathbf{E}_{\Phi}=0$,  i.e. on-shell, the conserved condition
	\begin{equation}\label{key}
	\mathrm{d} \bm{\omega} \left(\delta_{1} \Phi, \delta_{2} \Phi, \Phi\right) \approx 0
	\end{equation}
	is satisfied, where $\approx$ denotes the on-shell equality.
	
	When the $U(1)$ gauge field $A$ is present in $\mathbf{L}$, in addition to the diffeomorphism generated by a vector field $\xi$, the Lagrangian can also be gauge invariant under $A \rightarrow A+\mathrm{d} \lambda$ for an arbitrary scalar $\lambda$. So the general gauge transformation has the form $\delta_{\epsilon} \Phi=\left\{\mathcal{L}_{\xi} \Phi, \delta_{\lambda} \Phi\right\}$ where the generator $\epsilon$ is a combination of diffeomorphism+gauge transformation, $\epsilon \equiv\left\{\xi, \lambda\right\}$.
	
	For field-dependent transformations, i.e. $\delta \epsilon \neq 0$, the charge variations $\delta H_{\epsilon}(\Phi)$ associated with generators $\epsilon$ can be defined as
	\begin{align} \label{eqdh}
		\delta H_{\epsilon}(\Phi) & \equiv \Omega\left(\delta \Phi, \delta_{\epsilon} \Phi, \Phi\right)=\int_{\Sigma} \bm{\omega}\left(\delta \Phi, \delta_{\epsilon} \Phi, \Phi\right)  \nonumber \\  &\equiv \int_{\Sigma}\left(\delta^{[\Phi]} \mathbf{\Theta}\left(\delta_{\epsilon} \Phi, \Phi\right)-\delta_{\epsilon} \mathbf{\Theta}(\delta \Phi, \Phi)\right) \nonumber \\ &=\oint_{\partial \Sigma} \bm{k}_{\epsilon}(\delta \Phi, \Phi) ,
	\end{align}
    where
	\begin{equation}\label{key}
	\bm{\omega}\left(\delta \Phi, \delta_{\epsilon} \Phi, \Phi\right) \approx \mathrm{d} \bm{k}_{\epsilon}(\delta \Phi, \Phi)
	\end{equation}
	with
	\begin{equation}\label{k expression}
	\bm{k}_{\epsilon}(\delta \Phi, \Phi)=\delta \mathbf{Q}_{\epsilon}-\xi \cdot \mathbf{\Theta}(\delta \Phi, \Phi) ,
	\end{equation}
	in which $\mathbf{Q}_{\epsilon}$ is the Noether-Wald charge density defined by
	\begin{equation}\label{NW charge density}
	\mathrm{d} \mathbf{Q}_{\epsilon} \equiv \mathbf{\Theta} \left(\delta_{\epsilon} \Phi, \Phi\right)-\xi \cdot \mathbf{L} .
	\end{equation}
	In Eq.(\ref{eqdh}), $\delta^{[\Phi]}$ denotes that $\delta$ acts only on $\Phi$ and not on the $\epsilon$ inside $\mathbf{\Theta}$.
	
	If $\delta H_{\epsilon}(\Phi)$ is well-defined and integrable, we can find the conserved charge $H_{\epsilon}$. The integrability condition is basically $\left(\delta_{1} \delta_{2}-\delta_{2} \delta_{1}\right) H_{\epsilon}(\Phi)=0$, which is equivalent to \cite{Hajian:2015xlp,Compere:2015knw}
	\begin{equation}\label{intcondition}
	\oint_{\partial \Sigma}\left(\xi \cdot \bm{\omega}\left(\delta_{1} \Phi, \delta_{2} \Phi, \Phi\right)+\bm{k}_{\delta_{1} \epsilon}\left(\delta_{2} \Phi, \Phi\right)-\bm{k}_{\delta_{2} \epsilon}\left(\delta_{1} \Phi, \Phi\right)\right) \approx 0 .
	\end{equation}
	
	For guaranteeing the conservation of charge variation $\delta H_{\epsilon}$ and the independence of integration on $\Sigma$ and $\partial \Sigma$, the symplectic current form $\bm{\omega}$ must vanishes on-shell for a subclass of $\delta_{\epsilon} \Phi$'s, i.e.
	\begin{equation}\label{symplectic symmetry}
	\bm{\omega}\left(\delta \Phi, \delta_{\epsilon} \Phi, \Phi\right) \approx 0 .
	\end{equation}
	In this case the transformations generated by $\delta_{\epsilon} \Phi$ are called ``symplectic symmetries”. The family of $\epsilon$'s with this property can be divided into two sets: (1).``non-exact symmetry generators” denoted by $\chi$, for which $\delta_{\chi} \Phi \neq 0$ at least on one of the point of the phase space. (2).``exact symmetry generators” denoted by $\eta$, for which $\delta_{\eta} \Phi = 0$ all over the phase space.
	
	\textbf{Solution phase space method:} The SPSM is specification of CPSM to some specific manifolds and their tangent space. We consider the manifold $\hat {\mathcal{M}}$ to be composed of solutions $\hat {\Phi}=\hat{\Phi}\left(x ; p_{\alpha}\right)$, where $p_{\alpha}$ are a set of solution parameters. The parametric variations $\hat{\delta} \Phi$ are defined as
	 \begin{equation}\label{key}
	 \hat{\delta} \Phi \equiv \frac{\partial \hat{\Phi}}{\partial p_{\alpha}} \delta p_{\alpha} .
	 \end{equation}
	The corresponding symplectic $2$-form $\Omega$ can be denoted as $\hat\Omega$. The solution phase space is denoted by $( \hat{\mathcal{M}},\hat{\Omega})$, which is a submanifold of $( \mathcal{M},\Omega)$ when $\{\Phi \}$ limit to space $\{\hat\Phi \}$.	Charge variations associated with generators $\epsilon$ would be taken as
	\begin{equation}\label{dhintgration}
	\hat{\delta} H_{\epsilon}=\oint_{\partial \Sigma} \bm{k}_{\epsilon}(\hat{\delta} \Phi, \hat{\Phi}).
	\end{equation}
    The integrability condition (\ref{intcondition}) for parametric variations is
	\begin{equation}\label{intcondition2}
	\oint_{\partial \Sigma}(\xi \cdot \bm{\omega}(\hat{\delta}_{1} \Phi, \hat{\delta}_{2} \Phi, \hat{\Phi})+\bm{k}_{\hat{\delta}_{1} \epsilon}(\hat{\delta}_{2} \Phi, \hat{\Phi})-\bm{k}_{\hat{\delta}_{2} \epsilon}(\hat{\delta}_{1} \Phi, \hat{\Phi})) \approx 0 .
	\end{equation}
	
	When the integration of Eq.(\ref{dhintgration}) is well-defined and integrable over parameters $p_{\alpha}$, the conserved charges $H_{\epsilon}\left(p_{\alpha}\right)$ can be calculated by
	\begin{equation}\label{Hexpression}
	H_{\epsilon}[\hat{\Phi} (p_{\alpha})]=\int_{\bar{p}}^{p} \hat{\delta} H_{\epsilon}+H_{\epsilon}[\bar{\Phi} (\bar{p}_{\alpha})] ,
	\end{equation}
	where $H_{\epsilon}[\bar{\Phi} (\bar{p}_{\alpha})]$ is the reference point (i.e. constant of integration) for the $H_{\epsilon}$ defined on some specific reference field configuration $\bar{\Phi} (x^{\mu}; \bar{p}_{\alpha})$. In Sec.4 we will use Eq.(\ref{Hexpression}) to calculate conserved charges and derive the first laws of Einstein-aether black hole solutions associated with exact symmetries $\hat{\delta}_{\eta} \Phi=0$.

\section{Einstein-aether-Maxwell theory}

    Let us consider the Einstein-aether-Maxwell theory with cosmological constant $\Lambda$. The dynamical fields $\Phi$ would be the metric $g_{\mu \nu}$, aether field $u^\mu$, and Abelian 1-form gauge field $A$ governed by the Lagrangian \cite{Ding:2015kba}
    \begin{equation}\label{Lagrangian}
    L=\frac{1}{16 \pi G_{\mathrm{ae}}}\left(R-2 \Lambda+L_{\mathrm{ae}}-\frac{1}{4}\mathcal{N} F_{\mu \nu} F^{\mu \nu}\right) ,
    \end{equation}
    where $R$ is the Ricci scalar, $F=dA$ is the electromagnetic field strength, and $\mathcal{N}$ is a constant. The aether Lagrangian $L_{\mathrm{ae}}$ is given by
    \begin{equation}\label{key}
    L_{\mathrm{ae}}=-Z^{\mu \nu}{}_{\rho \sigma} \nabla_{\mu} u^{\rho} \nabla_{\nu} u^{\sigma}+\lambda\left(u^{2}+1\right) ,
    \end{equation}
    where $\lambda$ is a Lagrange multiplier. The tensor $Z^{\mu \nu}{}_{\rho \sigma}$ is defined as
    \begin{equation}\label{key}
    Z^{\mu \nu}{}_{\rho \sigma}=c_{1} g^{\mu \nu} g_{\rho \sigma}+c_{2} {\delta^\mu}_{\rho} {\delta^\nu}_{\sigma}+c_{3} {\delta^\mu}_{\sigma} {\delta^\nu}_{\rho}-c_{4} u^{\mu} u^{\nu} g_{\rho \sigma} ,
    \end{equation}
    in which $c_{i}(i=1,2,3,4)$ are coupling constants satisfying the following constraints \cite{Jacobson:2004ts}
    \begin{align}
    	 0 & \leq c_{13}<1, \\ \nonumber 0 & \leq c_{14}<2, \\ \nonumber 2+c_{123} &+2 c_{2}>0,
    \end{align}
    with $c_{i j} \equiv c_{i}+c_{j}$ and $c_{i j k} \equiv c_{i}+c_{j}+c_{k}$. The more stringent theoretical and observational constraints on $c_i$ are studied extensively \cite{Jacobson:2007fh}, through the ultra-high energy cosmic rays \cite{Elliott:2005va}, the solar system tests \cite{Eling:2003rd,Graesser:2005bg}, the binary pulsars \cite{Foster:2007gr,Yagi:2013ava}, and with the data of the gravitational wave events GW170817 and GRB170817A \cite{Gong:2018cgj,Oost:2018tcv}, and also using the data of the first black hole image \cite{Zhu:2019ura}. The constant $G_{\mathrm{ae}}$ is related to Newton's gravitational constant $G_{N}$ by $G_{\mathrm{ae}}=(1-c_{14}/2)G_{N}$. In what follows we will drop the subscript “$\mathrm{ae}$” for simpler notation.

    The Lagrangian $n$-form is the Hodge dual of $L$,
    \begin{equation}\label{key}
    \mathbf{L}=\star{L}=L\, \bm{\epsilon} ,
    \end{equation}
        where $ \bm{\epsilon}=\sqrt{-g}\,d^nx=\frac{\sqrt{-g}}{n!} \varepsilon_{\mu_{1} \cdots\mu_{n}} d x^{\mu_{1}} \wedge \cdots \wedge d x^{\mu_{n}} $ is the $n$-dimensional volum form, $g$ is the determinant of the metric $g_{\mu \nu}$, and $\varepsilon_{\mu_{1} \cdots\mu_{n}}$ is the Levi-Civita symbol.
        Variations of the Lagriangian $n$-form $\mathbf{L}$ with respect to $g_{\mu \nu}$, $u^{\mu}$, $A_{\mu}$, and $\lambda$
        \begin{equation}\label{key}
        	\delta \mathbf{L}(\Phi)=\frac{1}{16 \pi G}(\mathrm{E}^{\mu \nu}_{g} \delta g_{\mu \nu}+ \mathrm{Æ}_{\mu} \delta u^{\mu}+\mathrm{E}^{\nu}_{A} \delta A_{\nu}+\mathrm{E}_{\lambda} \delta \lambda)\bm{\epsilon} +\mathrm{d} \mathbf{\Theta}(\delta \Phi, \Phi)
        \end{equation}
       give the following EOMs
        \begin{align}
        \mathrm{E}^{\mu \nu}_{g}=&-(R^{\mu \nu}-\frac{1}{2} R g^{\mu \nu}+\Lambda g^{\mu \nu})  \nonumber \\  &+\frac{1}{2} L_{\mathrm{ae}} g^{\mu \nu}+c_{1}(\nabla^{\mu} u_{\rho} \nabla^{\nu}  u^{\rho}-\nabla_{\rho} u^{\mu} \nabla^{\rho} u^{\nu})+c_{4} u^{\rho}\nabla_{\rho} u^{\mu} u^{\sigma}\nabla_{\sigma}u^{\nu} \nonumber \\ &+\lambda u^{\mu}u^{\nu}+\nabla_{\rho}X^{\rho \mu \nu}   +\frac{1}{2}\mathcal{N}[{F^{\mu}}_{\rho}F^{\nu \rho}-\frac{1}{4}g^{\mu \nu}F_{\alpha \beta}F^{\alpha \beta}]=0 \label{Eg},\\
        \mathrm{Æ}_{\mu}=&2(\lambda u_{\mu}+c_{4}u^{\rho}\nabla_{\rho}u^{\sigma}\nabla_{\mu}u_{\sigma}+\nabla_{\rho}{Y^{\rho}}_{\mu})=0, \label{Eu} \\
        \mathrm{E}^{\nu}_{A}=&\nabla_{\mu}(\mathcal{N} F^{\mu \nu})=0 \label{Ee}, \\
        \mathrm{E}_{\lambda}=&u^2+1=0, \label{Elambda}
        \end{align}
        where
        \begin{align}
        {Y{^\mu}}_{\nu}=&Z^{\mu \rho}{}_{\nu \sigma}\nabla_{\rho}u^{\sigma},  \nonumber \\
        {X^{\sigma}}_{\mu \nu}=&{Y^{\sigma}}_{(\mu}u_{\nu )}-u_{(\mu}{Y_{\nu )}}^{\sigma}+u^{\sigma}Y_{(\mu \nu)}.
        \end{align}
        From (\ref{Eu}) and (\ref{Elambda}), we can get the Lagrange multiplier
        \begin{equation}\label{key}
        \lambda=c_{4} a^{2}+u^{\mu}\nabla_{\rho}{Y^{\rho}}_{\mu},
        \end{equation}
        where $a^{\rho}=u^{\sigma}\nabla_{\sigma}u^{\rho}$.

        The surface $n-1$-form $\mathbf{\Theta}(\delta \Phi, \Phi)$ is given by
        \begin{equation}\label{key}
        \mathbf{\Theta}(\delta \Phi, \Phi)=\Theta^{\mu}(\delta \Phi, \Phi) \sqrt{-g}(d^{n-1}x)_{\mu} ,
        \end{equation}
        in which
        \begin{equation}\label{key}
        \Theta^{\mu}(\delta \Phi, \Phi)=\frac{1}{16 \pi G}[2 \nabla^{[\sigma } {h^{\mu ]}}_{\sigma}-X^{\mu \alpha \beta} h_{\alpha \beta}-2 {Y^{\mu}}_{\nu} \delta u^{\nu}-\mathcal{N} F^{\mu \nu}\delta A_{\nu} ],
        \end{equation}
        where and in what follows $h_{\mu \nu }=\delta g_{\mu \nu },\,\,\,\,h^{\mu \nu }= g^{\mu \alpha } g^{\nu \beta }\delta g_{\alpha \beta }=-\delta g^{\mu \nu },\,\,\,\,h=g^{\mu \nu }\delta g_{\mu \nu }$, and $\left(d^{n-p} x\right)_{\mu_{1} \cdots \mu_{p}}=\frac{1}{p !(n-p) !} \varepsilon_{\mu_{1} \cdots \mu_{p} \nu_{p+1} \cdots \nu_{n}} d x^{\nu_{p+1}} \wedge \cdots \wedge d x^{\nu_{n}}$.

        For the generator $\epsilon=\{\xi,\lambda \}$, the transformations are
        \begin{align}
    	\delta_{\epsilon} g_{\mu \nu}&=\mathcal{L}_{\xi} g_{\mu \nu}=2 \nabla_{(\mu} \xi_{\nu)}, \\ \nonumber
    	\delta_{\epsilon} A_{\nu}&=\mathcal{L}_{\xi} A_{\nu}+\nabla_{\nu} \lambda=\xi^{\sigma} F_{\sigma \nu}+\nabla_{\nu}\left(A_{\sigma} \xi^{\sigma}+\lambda\right), \\ \nonumber
    	\delta_{\epsilon} u^{\mu}&=\mathcal{L}_{\xi} u^{\mu}=\xi^{\nu}\nabla_{\nu} u^{\mu}-u^{\nu}\nabla_{\nu}\xi^{\mu},
        \end{align}
        where $\mathcal{L}_{\xi} $ is the Lie derivative along the vector field $\xi$. Hence, from Eq.(\ref{NW charge density}) we can get the Noether-Wald charge density
        \begin{equation}\label{key}
        \mathbf{Q}_{\epsilon}=\mathrm{Q}^{\mu \nu}_{\epsilon}\sqrt{-g}(d^{n-2}x)_{\mu \nu} ,
        \end{equation}
        where
        \begin{equation}\label{key}
        \mathrm{Q}^{\mu \nu}_{\epsilon}=-\frac{1}{16 \pi G}[2 \nabla^{[\mu} \xi^{\nu]}+2 \xi_{\sigma}(Y^{[\mu \nu]} u^{\sigma}+u^{[\mu}Y^{\nu] \sigma}-Y^{\sigma [\mu} u^{\nu]})+\mathcal{N} F^{\mu \nu}(A_{\sigma}\xi^{\sigma}+\lambda)] .
        \end{equation}
        With the expressions above and using Eq.(\ref{k expression}) we get the surface charge
        \begin{equation}\label{surface charge expression}
        \bm{k}_{\epsilon}(\delta \Phi,\Phi)=\sqrt{-g}k^{\mu \nu}_{\epsilon}(d^{n-2}x)_{\mu \nu} \, ,
        \end{equation}
        where
        \begin{equation}\label{key}
        k^{\mu \nu}_{\epsilon}(\delta\Phi,\Phi)=k^{\mu \nu}_{\epsilon}(\delta g,g)+k^{\mu \nu}_{\epsilon}(\delta A,A)+k^{\mu \nu}_{\epsilon}(\delta u,u)
        \end{equation}
        with
        \begin{align}
        	k^{\mu \nu}_{\epsilon}(\delta g, g)=&-\frac{1}{8 \pi G} \bigg[\frac{1}{2} h \nabla^{[ \mu} \xi^{\nu ]} -h^{\sigma [\mu} \nabla_{\sigma} \xi^{\nu]}+\xi_{\sigma} \nabla^{[\mu} h^{\nu] \sigma}-\xi^{[\mu} \nabla_{\sigma} h^{\nu] \sigma}+\xi^{[\mu} \nabla^{\nu]} h \bigg] , \nonumber \\
        	k^{\mu \nu}_{\epsilon}(\delta A,A)=&-\frac{1}{16 \pi G} \bigg[\big( \frac{1}{2}  h \mathcal{N} F^{\mu \nu}+\mathcal{N}\delta F^{\mu \nu}\big) \big(A_{\sigma}\xi^{\sigma}+\lambda \big)+\mathcal{N}F^{\mu \nu} \delta A_{\sigma}\xi^{\sigma} -2 \mathcal{N}\xi^{[\nu}F^{\mu] \sigma}\delta A_{\sigma} \bigg] , \nonumber \\
        	k^{\mu \nu}_{\epsilon}(\delta u,u)=&-\frac{1}{8 \pi G} \bigg[\big( \frac{1}{2} h \xi_{\sigma}+h_{\sigma \rho }\xi^{\rho}\big) \big(Y^{[\mu \nu]} u^{\sigma}+u^{[\mu} Y^{\nu] \sigma}-Y^{\sigma [\mu} u^{\nu ]} \big) \nonumber \\
        	&+\xi_{\sigma} \big(Y^{[\mu \nu]} \delta u^{\sigma}+\delta u^{[\mu} Y^{\nu] \sigma}-Y^{\sigma [\mu} \delta u^{\nu ]}+\delta Y^{[\mu \nu]} u^{\sigma}+u^{[\mu} \delta Y^{\nu] \sigma}-\delta Y^{\sigma [\mu} u^{\nu ]} \big)  \nonumber \\
        	&-\xi^{[\nu}X^{\mu] \alpha \beta} h_{\alpha \beta}-2 \xi^{[\nu}{Y^{\mu]}}_{\sigma}\delta u^{\sigma} \bigg] ,
        \end{align}
        in which
        \begin{align}\label{key35}
        	\delta Y^{\mu \nu}=& \delta Z^{\mu}{} _{\rho}{}^{\nu}{}_{\sigma}  \nabla^{\rho} u^{\sigma}+ Z^{\mu}{} _{\rho}{}^{\nu}{}_{\sigma}  \bigg[-h^{\lambda \rho} \nabla_{\lambda} u^{\sigma}+\nabla^{\rho} \delta u^{\sigma}+u_{\lambda} \nabla^{[\rho} h^{\sigma] \lambda}+\frac{1}{2} u_{\lambda} \nabla^{\lambda} h^{\rho \sigma}  \bigg] ,  \nonumber \\
        	\delta Z^{\mu \nu}{}_{\rho \sigma}=& -c_{1} h^{\mu \nu} g_{\rho \sigma}+c_{1}  g^{\mu \nu} h_{\rho \sigma}-c_{4} u^{\mu} u^{\nu} h_{\rho \sigma}-c_{4} g_{\rho \sigma}(\delta u^{\mu} u^{\nu}+u^{\mu} \delta u^{\nu}) .
        \end{align}

\section{Conserved charges and first laws for Einstein-aether black holes}

        Having the result of the surface charge at hand, in this section we will calculate the conserved charges and derive the first laws for some examples of exact Einstein-aether black hole solutions both at KH and UH.

\subsection{ 3-dimensional static charged quasi-BTZ black hole}

        When $\mathcal{N}=4$ in Lagrangian (\ref{Lagrangian}), the 3-dimensional static charged quasi-BTZ black hole in branch $c_{14}=0, c_{123}\neq0$ is given by \cite{Ding:2016wcf}
        \begin{align}
        	ds^{2}=&-f(r) dt^{2}+\frac{dr^{2}}{f(r)}+r^{2}d \varphi^{2}, \nonumber \\
        	f(r)=&-m+(\bar\Lambda-{\Lambda^{\prime}}^{2}) r^{2}-\frac{q^{2}}{2} \log\frac{r}{l_{\mathrm{eff}}} , \nonumber \\
        	\alpha(r)=&\frac{1}{\sqrt{-m+\bar\Lambda r^2-\frac{q^2}{2}  \log \frac{r}{l_{\mathrm{eff}}}}+r \Lambda^{\prime}} ,\nonumber \\
        	\beta (r)=&-r \Lambda^{\prime}, \nonumber \\
        	e(r)=&-\sqrt{-m+\bar{\Lambda} r^{2}-\frac{q^{2}}{2} \log \frac{r}{l_{\mathrm{eff}}}}	,\nonumber \\
        	\hat{A}=&-\frac{q}{2} \log\frac{r}{l_{\mathrm{eff}}} \, \mathrm{d}t, \nonumber \\
        	\hat{u}^{\mu}=&(\alpha(r)-\frac{\beta(r)}{f(r)}, \beta(r), 0 ) ,
        \end{align}
        where $1 / l_{\mathrm{eff}}=\sqrt{\bar{\Lambda}}$, $\bar{\Lambda}=-\Lambda+\left(1+c_{123}+c_{2}\right) {\Lambda^{\prime}}^{2}$, and $\Lambda^{\prime}$ is a constant. Here and in the following examples B and C we denote $e(r)\equiv \hat{u}^{\mu} \zeta_{\mu} $ as an intermediate quantity to obtain $\alpha(r)$ when using the tetrad formalism to obtain the black hole solutions \cite{Ding:2016wcf}, in which $\zeta^{\mu}=(1,0,0)$ is a Killing vector. We will use it to obtain the radius of UH.

        In GR, the surface gravity and the electric potential at KH are usually defined by \cite{Carroll:2004st}
        \begin{align}
        \kappa_{{}_{\mathrm{KH}}}=&\sqrt{-\frac{1}{2}(\nabla_{\mu}\zeta_{\nu})(\nabla^{\mu}\zeta^{\nu})}|_{r_{{}_{\mathrm{KH}}}}\label{KHsg},  \\
        \Phi_{{}_{\mathrm{KH}}}=&\zeta_{\mu}A^{\mu}	|_{r_{{}_{\mathrm{KH}}}}\label{KHep}.
        \end{align}
        But at the UH in Einstein-aether theory, when one considers the peeling behavior of particles moving at any speed, the surface gravity and the electric potential are defined by \cite{Ding:2016wcf,Ding:2015kba,Cropp:2013sea}
        \begin{align}
        \kappa_{{}_{\mathrm{UH}}}=&\frac{1}{2}\nabla_{u}(u \cdot \zeta)|_{r_{{}_{\mathrm{UH}}}}\label{UHsg},  \\
        \Phi_{{}_{\mathrm{UH}}}=&\zeta_{\mu}A^{\mu}	|_{r_{{}_{\mathrm{UH}}}}\label{UHep},
        \end{align}
        where $u$ is the aether vector field and $\zeta$ is a Killing vector.

        Since the integration in Eq.(\ref{dhintgration}) is independent of the choice of integration surface $\partial \Sigma$ which surrounding the singularity of the black hole, we take $\partial \Sigma$ to be the circle of constant $(t,r)$ for simplicity and take the limit $r \rightarrow \infty$. Then, conserved charge variations for an exact symmetry $\eta$ would be
        \begin{equation}\label{deltaH}
        \hat{\delta} H_{\eta}=\oint_{\partial \Sigma} \bm{k}_{\eta}(\hat{\delta} \Phi, \hat{\Phi})=\int_{0}^{2 \pi} \lim_{r\to \infty}\sqrt{-\hat{g}} \, k_{\eta}^{t r}(\hat{\delta} \Phi, \hat{\Phi}) \mathrm{d} \varphi,
        \end{equation}
        where $k_{\eta}^{t r}$ is the $tr$ component of $k_{\eta}^{\mu \nu}$. The dynamical fields are $\hat\Phi=(\hat g_{\mu \nu}, \hat u^{\mu}, \hat A_{\mu})$, parametrized by $p_{\alpha}=\{m, q \}$. Inserting the parametric variations
        \begin{align}\label{parametric variations}
        	\hat{\delta} g_{\mu \nu}=&\frac{\partial \hat{g}_{\mu \nu}}{\partial m} \delta m+\frac{\partial \hat{g}_{\mu \nu}}{\partial q} \delta q ,\nonumber \\
        	\hat{\delta} u^{\mu}=&\frac{\partial \hat{u}^{\mu}}{\partial m} \delta m+\frac{\partial \hat{u}^{\mu}}{\partial q} \delta q , \nonumber \\
        	\hat{\delta} A_{\mu}=&\frac{\partial \hat{A}_{\mu}}{\partial m} \delta m+\frac{\partial \hat{A}_{\mu}}{\partial q} \delta q
        \end{align}
        in Eq.(\ref{deltaH}), one can calculate the conserved charges.

        \textbf{Mass:} Choosing the exact symmetry $\eta_{{}_{M}}=\{ \partial_{t}, 0 \}$ and substituting the surface charge expressions (\ref{surface charge expression})-(\ref{key35}) and the parametric variations (\ref{parametric variations}) into Eq.(\ref{deltaH}), we get
        \begin{equation}\label{key}
        \hat{\delta} M=\hat{\delta} H_{\eta_{{}_{\mathrm{M}}}}=\frac{\delta m}{8 G}  \quad \Rightarrow \quad M=\frac{m}{8 G} .
        \end{equation}
        \textbf{Electric charge:} Choosing the exact symmetry $\eta_{{}_{Q}}=\{0,1\}$ and by the similar procedure as calculating the mass, we obtain the electric charge as
        \begin{equation}\label{key}
        \hat{\delta} Q=\hat{\delta} H_{\eta_{{}_{\mathrm{Q}}}}=\frac{\delta q}{4 G}  \quad \Rightarrow \quad Q=\frac{q}{4 G} .
        \end{equation}
        \textbf{Entropy of KH:} For the KH of quasi-BTZ black hole, to satisfy the integrability condition (\ref{intcondition2}) (for more details see Appendix B of Ref.\cite{Hajian:2015xlp}), surface gravity, temperature, and electric potential are defined as
        \begin{equation}\label{BTZKH}
         \kappa_{{}_{\mathrm{KH}}}=r_{{}_{\mathrm{KH}}} (\bar\Lambda-{\Lambda^{\prime}}^{2})-\frac{q^{2}}{4 r_{{}_{\mathrm{KH}}}}, \quad T_{{}_{\mathrm{KH}}}=\frac{\kappa_{{}_{\mathrm{KH}}}}{2 \pi}, \quad \Phi_{{}_{\mathrm{KH}}}=-\frac{q}{2} \log\frac{r_{{}_{\mathrm{KH}}}}{l_{\mathrm{eff}}} ,
        \end{equation}
        where $r_{{}_{\mathrm{KH}}}$ denotes the radius of KH (i.e. $f(r_{{}_{\mathrm{KH}}})=0$). With the choice of the exact symmetry generator $\eta_{{}_{\mathrm{KH}}}=\frac{2 \pi}{\kappa_{{}_{\mathrm{KH}}}} \{\zeta_{{}_{\mathrm{KH}}}, -\Phi_{{}_{\mathrm{KH}}} \}$, in which $\zeta_{{}_{\mathrm{KH}}}=\partial_{t}$ is a time-like Killing vector, the corresponding entropy variation is given by
        \begin{equation}\label{key}
        \hat{\delta} S_{{}_{\mathrm{KH}}}=\hat{\delta} H_{\eta_{{}_{\mathrm{KH}}}}=\frac{2 \pi}{\kappa_{{}_{\mathrm{KH}}}}\left(\frac{1}{8 G} \delta m -\frac{\Phi_{{}_{\mathrm{KH}}}}{4 G} \delta q\right) .
        \end{equation}
        By the relations
        \begin{equation}\label{key}
        \frac{\partial r_{{}_{\mathrm{KH}}}}{\partial m}=\frac{1}{2 r_{{}_{\mathrm{KH}}} (\bar\Lambda-{\Lambda^{\prime}}^{2})-\frac{q^{2}}{2 r_{{}_{\mathrm{KH}}}}}, \quad \frac{\partial r_{{}_{\mathrm{KH}}}}{\partial q}=\frac{q \log \frac{r_{{}_{\mathrm{KH}}}}{l_{\mathrm{eff}}} }{2 r_{{}_{\mathrm{KH}}} (\bar\Lambda-{\Lambda^{\prime}}^{2})-\frac{q^{2}}{2 r_{{}_{\mathrm{KH}}}}}  ,
        \end{equation}
        we get
        \begin{equation}\label{key}
        \hat{\delta} S_{{}_{\mathrm{KH}}}=\frac{2 \pi}{4 G}\left(\frac{\partial r_{{}_{\mathrm{KH}}}}{\partial m} \delta m+\frac{\partial r_{{}_{\mathrm{KH}}}}{\partial q} \delta q\right)=\hat{\delta}\left(\frac{2 \pi r_{{}_{\mathrm{KH}}}}{4 G}\right) \quad \Rightarrow \quad S_{{}_{\mathrm{KH}}}=\frac{2 \pi r_{{}_{\mathrm{KH}}}}{4 G}=\frac{\mathrm{A_{{}_{\mathrm{KH}}}}}{4 G} ,
        \end{equation}
        where $\mathrm{A_{{}_{\mathrm{KH}}}}$ is the the area (here the perimeter) of KH.

        The reference points for the charges above are chosen to vanish when $m=q=0$.  \\
         \textbf{First law at KH:} With the decomposition
        \begin{equation}\label{key}
        \eta_{{}_{\mathrm{KH}}}=\frac{1}{T_{{}_{\mathrm{KH}}}}\left(\eta_{{}_{\mathrm{M}}}-\Phi_{{}_{\mathrm{KH}}} \eta_{{}_{\mathrm{Q}}}\right)
        \end{equation}
        and the linearity of $\hat\delta H_{\eta} $ in $\eta$, the first law at KH follows as
        \begin{equation}\label{3d first law}
         \delta S_{{}_{\mathrm{KH}}}=\frac{1}{T_{{}_{\mathrm{KH}}}}\left(\delta M-\Phi_{{}_{\mathrm{KH}}} \delta Q\right) .
        \end{equation}
        \textbf{Entropy of UH:} For the UH of quasi-BTZ black hole, the integrability condition (\ref{intcondition2}) requires the surface gravity, temperature, and electric potential be defined as
        \begin{equation}\label{BTZUH}
        \kappa_{{}_{\mathrm{UH}}}=\frac{\Lambda^{\prime}}{\sqrt{2}}\frac{r_{{}_{\mathrm{UH}}}}{l_{\mathrm{eff}}}, \quad T_{{}_{\mathrm{UH}}}=\frac{\kappa_{{}_{\mathrm{UH}}}}{2 \pi}, \quad \Phi_{{}_{\mathrm{UH}}}=\frac{m}{2 q}-\frac{\Lambda^{\prime}}{2 \sqrt{2}}\frac{q}{2 \sqrt{\bar\Lambda}} ,
        \end{equation}
        respectively, where $r_{{}_{\mathrm{UH}}}=\frac{q}{2 \sqrt{\bar\Lambda}}$ denotes the radius of UH (i.e. $e^2(r_{{}_{\mathrm{UH}}})=0,\, \frac{d e^2(r)}{d r} |_{r_{{}_{\mathrm{UH}}}}=0$). With the symmetry generator chosen as $\eta_{{}_{\mathrm{UH}}}=\frac{2 \pi}{\kappa_{{}_{\mathrm{UH}}}} \{\zeta_{{}_{\mathrm{UH}}}, -\Phi_{{}_{\mathrm{UH}}} \}$, in which $\zeta_{{}_{\mathrm{UH}}}=\partial_{t}$, the corresponding entropy variation is given by
        \begin{equation}\label{key}
        \hat{\delta} S_{{}_{\mathrm{UH}}}=\hat{\delta} H_{\eta_{{}_{\mathrm{UH}}}}=\frac{2 \pi}{\kappa_{{}_{\mathrm{UH}}}}\left(\frac{1}{8 G} \delta m -\frac{\Phi_{{}_{\mathrm{UH}}}}{4 G} \delta q\right)=\frac{2 \pi}{4 G} \delta \left (r_{{}_{\mathrm{UH}}}+\frac{\sqrt{2}}{\Lambda^{\prime}} \frac{m}{q}\right ).
        \end{equation}
        Therefore,
        \begin{equation}\label{key}
         S_{{}_{\mathrm{UH}}}=\frac{2 \pi(r_{{}_{\mathrm{UH}}}+\frac{\sqrt{2} m}{\Lambda^{\prime} q})}{4 G}=\frac{\mathrm{A_{{}_{\mathrm{UH}}}}}{4 G}+\frac{\sqrt{2} m} {4 G \Lambda^{\prime} q},
        \end{equation}
        where $\mathrm{A_{{}_{\mathrm{UH}}}}$ is the the area of UH. The reference point for the entropy $S_{{}_{\mathrm{UH}}}$ is chosen to vanish when $m=q=0$.   \\
        \textbf{First law at UH:} At UH, with the similar considerations of the decomposition $\eta_{{}_{\mathrm{UH}}}$ and the linearity of $\hat\delta H_{\eta} $ as those at KH, one can easily verify the first law as
        \begin{equation}\label{key}
        \delta S_{{}_{\mathrm{UH}}}=\frac{1}{T_{{}_{\mathrm{UH}}}}\left(\delta M-\Phi_{{}_{\mathrm{UH}}} \delta Q\right) .
        \end{equation}

        In Eq.(\ref{BTZKH}), the definitions of the surface gravity $\kappa_{{}_{\mathrm{KH}}}$ and the electric potential $\Phi_{{}_{\mathrm{KH}}}$ satisfy the integrability condition and are equivalent to the GR definitions (\ref{KHsg}) and (\ref{KHep}). The entropy is exactly one quarter of the area of KH, and the first law is satisfied exactly. At UH, to satisfy the integrability condition we have redefined the surface gravity and electric potential. The surface gravity is agree with the definition (\ref{UHsg}), but the electric potential is not proportional to (\ref{UHep}). We obtained the first law at UH, but the entropy has a corrected term in addition to the term proportional to the horizon area.

\subsection{ 4-dimensional static charged Einstein-aether black hole for $c_{14}=0, c_{123}\neq0$}

        With $\mathcal{N}=4$ in (\ref{Lagrangian}), the 4-dimensional static charged Einstein-aether black hole solution in branch $c_{14}=0, c_{123}\neq0$ is given by \cite{Ding:2015kba, Zhu:2019ura}
        \begin{align}
        	ds^{2}=&-f(r) dt^{2}+\frac{dr^{2}}{f(r)}+r^{2}(d \theta^{2}+\sin^{2} \theta d \varphi^{2}), \nonumber \\
        	f(r)=&1-\frac{2 m}{r}+\frac{q^{2}}{r^{2}}-c\frac{m^{4}}{r^{4}} , \nonumber \\
        	\alpha(r)=&\left[\sqrt{1-\frac{2 m}{r}+\frac{q^{2}}{r^{2}}-c^{\prime}\frac{m^{4}}{r^{4}}}+\bar{c}\left(\frac{2 m}{r}\right)^{2}\right]^{-1} ,\nonumber \\
        	\beta(r)=&-\bar{c}\left(\frac{2 m}{r}\right)^{2}, \nonumber \\
            e(r)=&-\sqrt{1-\frac{2 m}{r}+\frac{q^{2}}{r^{2}}-c^{\prime}\frac{m^{4}}{r^{4}}}	,\nonumber \\
        	\hat{A}=&-\frac{q}{r} \, \mathrm{d}t, \nonumber \\
        	\hat{u}^{\mu}=&(\alpha(r)-\frac{\beta(r)}{f(r)}, \beta(r), 0 ,0) ,
        \end{align}
        where $c=\frac{27 c_{13}}{16(1-c_{13})}$, $\bar{c}=\frac{3 \sqrt{3}}{16 \sqrt{1-c_{13}}}$, $c^{\prime}=-\frac{27}{16}$.

        With the integration surface $\partial \Sigma$ in Eq.(\ref{dhintgration}) taken to be the sphere of constant $(t,r)$, conserved charge variations for an exact symmetry $\eta$ would be
        \begin{equation}\label{deltaH41}
        \hat{\delta} H_{\eta}=\oint_{\partial \Sigma} \bm{k}_{\eta}(\hat{\delta} \Phi, \hat{\Phi})=\int_{0}^{2 \pi} \int_{0}^{\pi} \lim_{r\to \infty}\sqrt{-\hat{g}} \, k_{\eta}^{t r}(\hat{\delta} \Phi, \hat{\Phi}) \mathrm{d} \theta \, \mathrm{d} \varphi.
        \end{equation}
        The dynamical fields are still $\hat\Phi=(\hat g_{\mu \nu}, \hat u^{\mu}, \hat A_{\mu})$ parametrized by $p_{\alpha}=\{m, q \}$. Thus, the similar expressions of the parametric variations as Eq.(\ref{parametric variations}) can be inserted to give the conserved charges as follows.  \\
        \textbf{Mass:} The exact symmetry is still chosen as $\eta_{{}_{M}}=\{ \partial_{t}, 0 \}$ and the mass is resulted as
        \begin{equation}\label{key}
        \hat{\delta} M=\hat{\delta} H_{\eta_{{}_{\mathrm{M}}}}=\frac{\delta m}{ G}  \quad \Rightarrow \quad M=\frac{m}{ G} .
        \end{equation}
        \textbf{Electric charge:} Choosing the exact symmetry $\eta_{{}_{Q}}=\{0,-1\}$, in which the minus sign is to guarantee the positive electric charge,
        \begin{equation}\label{key}
        \hat{\delta} Q=\hat{\delta} H_{\eta_{{}_{\mathrm{Q}}}}=\frac{\delta q}{ G}  \quad \Rightarrow \quad Q=\frac{q}{ G} .
        \end{equation}
        \textbf{Entropy of KH:} The following definitions of the surface gravity, temperature, and electric potential for the KH of  4-dimensional static charged Einstein-aether black hole satisfy the integrability condition (\ref{intcondition2}),
        \begin{equation}\label{c140KH}
        \kappa_{{}_{\mathrm{KH}}}= \frac{2 r^{3}_{{}_{\mathrm{KH}}}-3 m r^{2}_{{}_{\mathrm{KH}}}+r_{{}_{\mathrm{KH}}} q^2}{ r_{{}_{\mathrm{KH}}} (r^{3}_{{}_{\mathrm{KH}}}+2 c m^{3})}, \quad T_{{}_{\mathrm{KH}}}=\frac{\kappa_{{}_{\mathrm{KH}}}}{2 \pi}, \quad \Phi_{{}_{\mathrm{KH}}}=\frac{r^{2}_{{}_{\mathrm{KH}}} q}{r^{3}_{{}_{\mathrm{KH}}}+2 c m^{3}} .
        \end{equation}
        With the exact symmetry generator $\eta_{{}_{\mathrm{KH}}}=\frac{2 \pi}{\kappa_{{}_{\mathrm{KH}}}} \{\zeta_{{}_{\mathrm{KH}}}, \Phi_{{}_{\mathrm{KH}}} \}$, in which $\zeta_{{}_{\mathrm{KH}}}=\partial_{t}$, the corresponding entropy variation is expressed as
        \begin{equation}\label{key}
        \hat{\delta} S_{{}_{\mathrm{KH}}}=\hat{\delta} H_{\eta_{{}_{\mathrm{KH}}}}=\frac{2 \pi}{\kappa_{{}_{\mathrm{KH}}}}\left(\frac{1}{ G} \delta m -\frac{\Phi_{{}_{\mathrm{KH}}}}{ G} \delta q\right) .
        \end{equation}
        With the relations
        \begin{equation}\label{key}
        \frac{\partial r_{{}_{\mathrm{KH}}}}{\partial m}=\frac{r^{3}_{{}_{\mathrm{KH}}}+2 c m^{3}}{2 r^{3}_{{}_{\mathrm{KH}}}-3 m r^{2}_{{}_{\mathrm{KH}}}+r_{{}_{\mathrm{KH}}} q^2}, \quad \frac{\partial r_{{}_{\mathrm{KH}}}}{\partial q}=\frac{-r^{2}_{{}_{\mathrm{KH}}} q}{2 r^{3}_{{}_{\mathrm{KH}}}-3 m r^{2}_{{}_{\mathrm{KH}}}+r_{{}_{\mathrm{KH}}} q^2} ,
        \end{equation}
        we have
        \begin{equation}\label{key}
        \hat{\delta} S_{{}_{\mathrm{KH}}}=\frac{\pi}{ G}\left(\frac{\partial r^{2}_{{}_{\mathrm{KH}}}}{\partial m} \delta m+\frac{\partial r^{2}_{{}_{\mathrm{KH}}}}{\partial q} \delta q\right)=\hat{\delta}\left(\frac{4 \pi r^{2}_{{}_{\mathrm{KH}}}}{4 G}\right) \quad \Rightarrow \quad S_{{}_{\mathrm{KH}}}=\frac{4 \pi r^{2}_{{}_{\mathrm{KH}}}}{4 G}=\frac{\mathrm{A_{{}_{\mathrm{KH}}}}}{4 G} .
        \end{equation}
        Still, we choose the reference points for the charges above to vanish when $m=q=0$.  \\
        \textbf{First law at KH:} Similar to the analysis of that in the previous example, the first law at KH of 4-dimensional static charged Einstein-aether black hole is easy to check, which has the same expression as Eq.(\ref{3d first law}). \\
        \textbf{Entropy of UH:} The surface gravity, temperature, and electric potential for the UH of 4-dimensional static charged Einstein-aether black hole can be defined as
        \begin{equation}\label{c140UH}
        \kappa_{{}_{\mathrm{UH}}}= \frac{2 r^{3}_{{}_{\mathrm{UH}}}-3 m r^{2}_{{}_{\mathrm{UH}}}+r_{{}_{\mathrm{UH}}} q^2}{ r_{{}_{\mathrm{UH}}} (r^{3}_{{}_{\mathrm{UH}}}+2 c^{\prime} m^{3})}, \quad T_{{}_{\mathrm{UH}}}=\frac{\kappa_{{}_{\mathrm{UH}}}}{2 \pi}, \quad \Phi_{{}_{\mathrm{UH}}}=\frac{r^{2}_{{}_{\mathrm{UH}}} q}{r^{3}_{{}_{\mathrm{UH}}}+2 c^{\prime} m^{3}} .
        \end{equation}
        Similar to the analysis of the entropy of KH for the same black hole, we can get the entropy of UH as
        \begin{equation}\label{key}
        S_{{}_{\mathrm{UH}}}=\frac{4 \pi r^{2}_{{}_{\mathrm{UH}}}}{4 G}=\frac{\mathrm{A_{{}_{\mathrm{UH}}}}}{4 G} .
        \end{equation}
        \textbf{First law at UH:} The first law at UH of 4-dimensional static charged Einstein-aether black hole is satisfied, which is easy to check.

        From the investigation above we can see that the entropies are exactly one quarter of the horizon areas, and the first laws of black hole thermodynamics are obtained both at KH and UH. In order to satisfy the integrability condition we have redefined the horizon quantities (\ref{c140KH}) and (\ref{c140UH}), which are not proportional to the definitions (\ref{KHsg}) and (\ref{KHep}). When $c_{13}=0$ this solution reduces to the Reissner-Nordstrom black hole, and our definitions (\ref{c140KH}) reduce to definitions (\ref{KHsg}) and (\ref{KHep}) at KH. This indicates that the contribution of aether vector field leads to the modification of the definitions of surface gravity and electric potential.

\subsection{ 4-dimensional static charged Einstein-aether black hole for $c_{14}\neq 0, c_{123}=0$}

        Still with $\mathcal{N}=4$ in (\ref{Lagrangian}), the 4-dimensional static charged Einstein-aether black hole solution in branch $c_{14}\neq0, c_{123}=0$ is given by \cite{Ding:2015kba}
        \begin{align}
        	ds^{2}=&-f(r) dt^{2}+\frac{dr^{2}}{f(r)}+r^{2}(d \theta^{2}+\sin^{2} \theta d \varphi^{2}), \nonumber \\
        	f(r)=&1-\frac{2 m}{r}+B_{1} \frac{q^{2}}{r^{2}}+B_{2} \frac{m^{2}}{r^{2}} , \nonumber \\
        	\alpha(r)=&\left[1-\left(\frac{1}{2}-B_{3}\right) \frac{2 m}{r}\right]^{-1} ,\nonumber \\
        	\beta(r)=&-B_{3} \frac{2 m}{r}, \nonumber \\
        	e(r)=&-1+\frac{m}{r}	,\nonumber \\
        	\hat{A}=&-\frac{q}{r} \, \mathrm{d}t, \nonumber \\
        	\hat{u}^{\mu}=&(\alpha(r)-\frac{\beta(r)}{f(r)}, \beta(r), 0 ,0) ,
        \end{align}
        where $B_{1}=\frac{1}{1-c_{13}}, \, B_{2}=-\frac{2 c_{13}-c_{14}}{2(1-c_{13})}, \, B_{3}=\frac{1}{2 \sqrt{1-c_{13} }}(1-\frac{c_{14}}{2}-\frac{q^{2}}{m^{2}})^{1 / 2}$.

        With the similar treatment as in the second example, we can get the following conserved charges. \\
        \textbf{Mass:} Choosing the exact symmetry $\eta_{{}_{M}}=\{ \partial_{t}, 0 \}$,
        \begin{equation}\label{key}
        \hat{\delta} M=\hat{\delta} H_{\eta_{{}_{\mathrm{M}}}}=\frac{2-c_{14}}{2 G} \delta m  \quad \Rightarrow \quad M=\frac{2-c_{14}}{2 G} m .
        \end{equation}
        \textbf{Electric charge:} Choosing the exact symmetry $\eta_{{}_{Q}}=\{0,-1\}$,
        \begin{equation}\label{key}
        \hat{\delta} Q=\hat{\delta} H_{\eta_{{}_{\mathrm{Q}}}}=\frac{\delta q}{ G}  \quad \Rightarrow \quad Q=\frac{q}{ G} .
        \end{equation}
        \textbf{Entropy of KH:} The surface gravity, temperature and electric potential are defined as
        \begin{equation}\label{key}
        \kappa_{{}_{\mathrm{KH}}}=(1-\frac{c_{14}}{2}) \cdot \frac{r_{{}_{\mathrm{KH}}}-m}{ r_{{}_{\mathrm{KH}}} (r_{{}_{\mathrm{KH}}}-B_{2} m)}, \quad T_{{}_{\mathrm{KH}}}=\frac{\kappa_{{}_{\mathrm{KH}}}}{2 \pi}, \quad \Phi_{{}_{\mathrm{KH}}}=(1-\frac{c_{14}}{2}) \cdot \frac{B_{1}q}{ r_{{}_{\mathrm{KH}}}-B_{2} m} ,
        \end{equation}
        respectively. The corresponding entropy variation is given by
        \begin{equation}\label{key}
        \hat{\delta} S_{{}_{\mathrm{KH}}}=\hat{\delta} H_{\eta_{{}_{\mathrm{KH}}}}=\frac{2 \pi}{\kappa_{{}_{\mathrm{KH}}}}\left(\frac{2-c_{14}}{2 G} \delta m -\frac{\Phi_{{}_{\mathrm{KH}}}}{ G} \delta q\right) .
        \end{equation}
        By the relations
        \begin{equation}\label{key}
        \frac{\partial r_{{}_{\mathrm{KH}}}}{\partial m}=\frac{r_{{}_{\mathrm{KH}}}-B_{2} m}{ r_{{}_{\mathrm{KH}}}- m}, \quad \frac{\partial r_{{}_{\mathrm{KH}}}}{\partial q}=\frac{-B_{1}q}{ r_{{}_{\mathrm{KH}}}- m} ,
        \end{equation}
        we can get the entropy of KH for 4-dimensional static charged Einstein-aether black hole with $c_{14}\neq 0$ and $c_{123}=0$, i.e. $S_{{}_{\mathrm{KH}}}=\frac{4 \pi r^{2}_{{}_{\mathrm{KH}}}}{4 G}=\frac{\mathrm{A_{{}_{\mathrm{KH}}}}}{4 G}$. \\
        \textbf{First law at KH:} The satisfaction of the first law at KH of this black hole solution is still easy to check. \\
        \textbf{Entropy of UH:} The surface gravity, temperature and electric potential of UH are defined as
        \begin{equation}\label{4DUH}
        \kappa_{{}_{\mathrm{UH}}}=(1-\frac{c_{14}}{2}) \cdot \frac{1}{ r_{{}_{\mathrm{UH}}}}, \quad T_{{}_{\mathrm{UH}}}=\frac{\kappa_{{}_{\mathrm{UH}}}}{2 \pi}, \quad \Phi_{{}_{\mathrm{UH}}}=0 .
        \end{equation}
        With the similar analysis as for the entropy of KH for the same black hole solution, we get the entropy of UH as $S_{{}_{\mathrm{UH}}}=\frac{4 \pi r^{2}_{{}_{\mathrm{UH}}}}{4 G}=\frac{\mathrm{A_{{}_{\mathrm{UH}}}}}{4 G}$. \\
        \textbf{First law at UH:} Since $\Phi_{{}_{\mathrm{UH}}}=0$, the first law at UH can be checked as
        \begin{equation}\label{key}
        \delta S_{{}_{\mathrm{UH}}}=\frac{1}{T_{{}_{\mathrm{UH}}}}\delta M .
        \end{equation}

        In this example, we have re-derived the total mass and electric charge using the SPSM, which agree with the results of the original paper \cite{Ding:2015kba}. However, our calculation has an advantage in that it does not need to consider the renormalization. The entropies are exactly one quarter of the horizon area both at KH and UH. At the UH the $\delta Q$ term is absent in the first law, which agrees with the result in \cite{Ding:2015kba} by using the Smarr method \cite{Smarr:1972kt}. And the surface gravity $\kappa_{{}_{\mathrm{UH}}}$ agrees with the result in \cite{Ding:2015kba} up to a factor $1/4$. The reason is that the $r_{{}_{\mathrm{UH}}}=m$ is not relevant to the electric charge, thus the electric potential can be set to zero.


\subsection{(2+1)-dimensional rotating asymptotically AdS black hole}

        (2+1)-dimensional neutral exact fully rotating black hole solutions of IR Hořava theory were found in \cite{Sotiriou:2014gna} in branch $c_{14}=0$. They are also solutions of aether theory, i.e. satisfying the EOM (\ref{Eg})-(\ref{Elambda}) when $\mathcal{N}=0$. For the sake of concreteness, in this subsection we will consider (2+1)-dimensional rotating asymptotically AdS black hole \cite{Sotiriou:2014gna,Pacilio:2017emh}
        \begin{align}
        	ds^{2}=&-f(r) dt^{2}+\frac{dr^{2}}{f(r)}+r^{2}(d \varphi+\Omega (r)dt)^{2}, \nonumber \\
        	f(r)=&-m+D_{1}  \frac{a^{2}}{r^{2}}+D_{2} \frac{m^{2}}{r^{2}}-\Lambda r^{2} , \nonumber \\
        	\Omega(r)=&-\frac{a}{2 r^{2}} ,\nonumber \\
        	\alpha(r)=&\frac{1}{2 r}\sqrt{-\frac{m^{2}}{(1-c_{13})\Lambda}-a^{2}} ,\nonumber \\
        	e(r)=&-\frac{m+2 r^{2} \Lambda}{2 \sqrt{-\Lambda} r}	,\nonumber \\
        	\hat{u}^{\mu} =&\left(-\frac{e(r)}{f(r)},-\alpha(r),\frac{e(r)\Omega(r)}{f(r)}\right) ,
        \end{align}
        where $D_{1}=1/4, \, D_{2}=\frac{c_{13}}{4(1-c_{13})\Lambda}, \, \Lambda=-1/l^{2}$, $l$ is AdS radius.

        For this kind of black holes, the conserved charge variations for an exact symmetry $\eta$ have the same expression as Eq.(\ref{deltaH}).
        The dynamical fields are $\hat\Phi=(\hat g_{\mu \nu}, \hat u^{\mu})$, parametrized by $p_{\alpha}=\{m, a \}$, and the parametric variations are given by
        \begin{align}\label{parametric variations3r }
        	\hat{\delta} g_{\mu \nu}=&\frac{\partial \hat{g}_{\mu \nu}}{\partial m} \delta m+\frac{\partial \hat{g}_{\mu \nu}}{\partial a} \delta a ,\nonumber \\
        	\hat{\delta} u^{\mu}=&\frac{\partial \hat{u}^{\mu}}{\partial m} \delta m+\frac{\partial \hat{u}^{\mu}}{\partial a} \delta a .
        	\end{align}
        Then, the conserved charges can be calculated as follows. \\
        \textbf{Mass:} Choosing the exact symmetry $\eta_{{}_{M}}=\{ \partial_{t}, 0 \}$,
        \begin{equation}\label{key}
        \hat{\delta} M=\hat{\delta} H_{\eta_{{}_{\mathrm{M}}}}=\frac{\delta m }{8 G}  \quad \Rightarrow \quad M=\frac{m}{8 G} .
        \end{equation}
        \textbf{Angular momentum:} Choosing the exact symmetry $\eta_{{}_{\mathrm{J}}}=\{-\partial_{\varphi},0\}$,
        \begin{equation}\label{key}
        \hat{\delta} J=\hat{\delta} H_{\eta_{{}_{\mathrm{J}}}}=(1-c_{13})\frac{\delta a}{ 8G}  \quad \Rightarrow \quad J=(1-c_{13})\frac{a}{ 8G} .
        \end{equation}
        \textbf{Entropy of KH:} For the KH of 3-dimensional rotating Einstein-aether black hole, to satisfy the integrability condition (\ref{intcondition2}), the surface gravity, temperature, and angular velocity are defined as
        \begin{equation}\label{3rKH}
        \kappa_{{}_{\mathrm{KH}}}=-\frac{r_{{}_{\mathrm{KH}}} m+2 \Lambda r^{3}_{{}_{\mathrm{KH}}}}{ r^{2}_{{}_{\mathrm{KH}}}-D_{2} m}, \quad T_{{}_{\mathrm{KH}}}=\frac{\kappa_{{}_{\mathrm{KH}}}}{2 \pi}, \quad \Omega_{{}_{\mathrm{KH}}}=\frac{1}{1-c_{13}} \cdot \frac{2 D_{1} a}{ r^{2}_{{}_{\mathrm{KH}}}-D_{2} m} .
        \end{equation}
       With the exact symmetry generator chosen as $\eta_{{}_{\mathrm{KH}}}=\frac{2 \pi}{\kappa_{{}_{\mathrm{KH}}}}\zeta_{{}_{\mathrm{KH}}}$, in which $\zeta_{{}_{\mathrm{KH}}}=\partial_{t}+\Omega_{{}_{\mathrm{KH}}}\partial_{\varphi}$, the corresponding entropy variation is given by
        \begin{equation}\label{key}
        \hat{\delta} S_{{}_{\mathrm{KH}}}=\hat{\delta} H_{\eta_{{}_{\mathrm{KH}}}}=\frac{2 \pi}{\kappa_{{}_{\mathrm{KH}}}}\left(\frac{1}{8 G} \delta m -\Omega_{{}_{\mathrm{KH}}} \frac{1-c_{13}}{8 G} \delta a\right) .
        \end{equation}
        With the relations
        \begin{equation}\label{key}
        \frac{\partial r_{{}_{\mathrm{KH}}}}{\partial m}=-\frac{r^{2}_{{}_{\mathrm{KH}}}-2 D_{2} m}{2 (r_{{}_{\mathrm{KH}}} m+2\Lambda r^{3}_{{}_{\mathrm{KH}}})}, \quad \frac{\partial r_{{}_{\mathrm{KH}}}}{\partial a}=\frac{ D_{1} a}{r_{{}_{\mathrm{KH}}} m+2\Lambda r^{3}_{{}_{\mathrm{KH}}}} ,
        \end{equation}
        we get
        \begin{equation}\label{key}
        \hat{\delta} S_{{}_{\mathrm{KH}}}=\frac{2 \pi}{4 G}\left(\frac{\partial r_{{}_{\mathrm{KH}}}}{\partial m} \delta m+\frac{\partial r_{{}_{\mathrm{KH}}}}{\partial a} \delta a\right)=\hat{\delta}\left(\frac{2 \pi r_{{}_{\mathrm{KH}}}}{4 G}\right) \quad \Rightarrow \quad S_{{}_{\mathrm{KH}}}=\frac{2 \pi r_{{}_{\mathrm{KH}}}}{4 G}=\frac{\mathrm{A_{{}_{\mathrm{KH}}}}}{4 G} .
        \end{equation}
       The reference points for the charges above are chosen to vanish when $m=a=0$.  \\
        \textbf{First law at KH:} With the decomposition
        \begin{equation}\label{key}
        \eta_{{}_{\mathrm{KH}}}=\frac{1}{T_{{}_{\mathrm{KH}}}}\left(\eta_{{}_{\mathrm{M}}}-\Omega_{{}_{\mathrm{KH}}} \eta_{{}_{\mathrm{J}}}\right)
        \end{equation}
        and the linearity of $\hat\delta H_{\eta} $ in $\eta$, the first law at KH follows as
        \begin{equation}\label{key}
        \delta S_{{}_{\mathrm{KH}}}=\frac{1}{T_{{}_{\mathrm{KH}}}}\left(\delta M-\Omega_{{}_{\mathrm{KH}}} \delta J\right) .
        \end{equation}
        \textbf{Entropy of UH:} The surface gravity, temperature and electric potential for the UH of 3-dimensional rotating Einstein-aether black hole can be defined as
        \begin{equation}\label{3rUH}
        \kappa_{{}_{\mathrm{UH}}}=-2 \Lambda r_{{}_{\mathrm{UH}}}, \quad T_{{}_{\mathrm{UH}}}=\frac{\kappa_{{}_{\mathrm{UH}}}}{2 \pi}, \quad \Omega_{{}_{\mathrm{UH}}}=\frac{8 \Lambda r_{{}_{\mathrm{UH}}} a}{1-c_{13}} ,
        \end{equation}
        where $r_{{}_{\mathrm{UH}}}=\sqrt{-m/2 \Lambda}$ denotes the radius of UH (i.e. $e(r_{{}_{\mathrm{UH}}})=0$). Choosing the exact symmetry generator $\eta_{{}_{\mathrm{UH}}}=\frac{2 \pi}{\kappa_{{}_{\mathrm{UH}}}}\zeta_{{}_{\mathrm{UH}}}$, in which $\zeta_{{}_{\mathrm{UH}}}=\partial_{t}+\Omega_{{}_{\mathrm{UH}}}\partial_{\varphi}$, the corresponding entropy variation is given by
        \begin{equation}\label{key}
        \hat{\delta} S_{{}_{\mathrm{UH}}}=\hat{\delta} H_{\eta_{{}_{\mathrm{UH}}}}=\frac{2 \pi}{\kappa_{{}_{\mathrm{UH}}}}\left(\frac{1}{8 G} \delta m -\Omega_{{}_{\mathrm{UH}}} \frac{1-c_{13}}{8 G} \delta a\right)=\frac{2 \pi}{4 G} \delta(r_{{}_{\mathrm{UH}}}+a^{2}).
        \end{equation}
        Therefore,
        \begin{equation}\label{key}
         S_{{}_{\mathrm{UH}}}=\frac{2 \pi (r_{{}_{\mathrm{UH}}}+a^{2})}{4 G}=\frac{\mathrm{A_{{}_{\mathrm{UH}}}}}{4 G}+\frac{\pi a^{2}}{2 G} .
        \end{equation}
        \textbf{First law at UH:} It is easy to check that the first law at UH $\delta S_{{}_{\mathrm{UH}}}=\frac{1}{T_{{}_{\mathrm{UH}}}}\left(\delta M-\Omega_{{}_{\mathrm{UH}}} \delta J\right)$ is satisfied with the decomposition $\eta_{{}_{\mathrm{UH}}}=\frac{1}{T_{{}_{\mathrm{UH}}}}\left(\eta_{{}_{\mathrm{M}}}-\Omega_{{}_{\mathrm{UH}}} \eta_{{}_{\mathrm{J}}}\right)$ and the linearity of $\hat\delta H_{\eta} $ in $\eta$.

        In our formulation the total mass and angular momentum agree with the results in \cite{Pacilio:2017emh}. At KH for satisfing the integrability condition we have redefined the surface gravity and the angular velocity, and the entropy is exactly one quarter of the horizon area. But at UH the entropy $S_{{}_{\mathrm{UH}}}$ has a corrected term in addition to the one proportional to horizon area.

 \section{conclusions and remarks}
        In this paper, we use the SPSM to investigate the thermodynamics of black holes in Einstein-aether-Maxwell Theory. We derive the first laws of thermodynamics and definitive entropy expressions at both KH and UH for some examples of exact black hole solutions, including 3-dimensional static charged quasi-BTZ black hole, two 4-dimensional static charged black holes and 3-dimensional rotating solution. We find that at KHs the entropies are exactly one quarter of the horizon area, but at UHs of 3-dimensional black holes the entropies are not proportional to the horizon area. From the expression for the surface charge we derived, we re-computed the total mass, electric charge and angular momentum for these black hole solutions, confirming the results obtained in the previous literature.

        In the SPSM, the surface of integration can be an arbitrary codimension-2 surface surrounding the singularity and need not be the horizon itself. Thus, by this freedom of the choice of integral surface, the divergence of aether field at KH does not affect the final results and the SPSM is still valid at KH. In our calculation, the conserved charges are automatically regular and need not be renormalized. The formulations present in this paper can be easily generalized to any dimensional black hole solutions in any generally covariant gravitational theories.

        In general, if a horizon has an entropy, the integrability condition should be satisfied so that we can get a definitive entropy expression. By the linearity of the generator in Hamiltonian variation, the first law of corresponding horizon can be derived. We must note that, in our framework if we use Eqs.(\ref{KHsg})-(\ref{UHep}) or the tunneling temperature defined in \cite{Berglund:2012fk,Ding:2015fyx,Cropp:2013sea,Michel:2015rsa} as the definitions of the horizon quantities (temperature, electric potential and angular velocity), the integrability condition (\ref{intcondition2}) can not be satisfied except for the example in subsection IV.A. Correspondingly, we can’t get the definitive entropy expressions and there are no first laws both at KH and UH. In general, our definitions of the horizon quantities are based on: (1) integrability condition; (2) the entropy is proportional to the horizon area as much as possible. Usually, to satisfy the above two conditions we can define the surface gravity by $\frac{1}{\kappa_{H}}=\frac{\alpha}{n-2}\frac{\partial r^{n-2}_{H}}{\partial m}$, in which $\alpha$ is a constant irrelevant to the spacetime coordinates and the solution parameters. In the Lorentz symmetric gravitational theories \cite{Hajian:2015xlp,Hajian:2016kxx,Ghodrati:2016vvf}, it (almost) agrees with the GR definitions (\ref{KHsg}) and (\ref{KHep}). To satisfy the integrability condition, in Einstein-aether theory we have redefined the horizon quantities, which are not proportional to the general definitions in GR and the ones defined by the tunneling method \cite{Berglund:2012fk,Ding:2015fyx,Cropp:2013sea,Michel:2015rsa}. Therefore, the remaining questions are which temperature is a useful definition and what the asymptotic observers see? These questions are yet to be investigated.

        On the other hand, from our new definitions of surface gravity (thereby the temperature) and electric potential, we can see the aether field has a contribution to them. Thus, we can infer that a meaningful black hole solution in the Einstein-aether-Maxwell theory may entail the coupling between Maxwell field and aether field. Our formulation suggests an alternative way to show the first laws of black hole thermodynamics in LV gravitational theories.


\acknowledgments
This work is supported partially by National Science Foundation of China grant No. 10671128 and the Key Project of Chinese Ministry of Education grant No. 211059.

\end{document}